\documentclass[]{aa}
\usepackage{times}
\usepackage[]{graphics}
\usepackage[]{psfig}

\begin{document}

   \thesaurus{12.07.1, 12.03.3, 11.17.3, 11.17.2, 03.20.1, 03.20.8}
   \title{Exploring the gravitationally lensed system HE 1104-1805: \\Near-IR
Spectroscopy\thanks{Based on observations collected with the ESO New
Technology Telescope (program 61.B-0413)}}

   \author{F. Courbin
          \inst{1}
	  \and
	  C. Lidman
	  \inst{2}
	  \and
	  G. Meylan
	  \inst{3}
          \and
          J.-P. Kneib
          \inst{4}
          \and
          P. Magain 
          \inst{5}
          }

   \offprints{F. Courbin}

   \institute{Universidad Cat\'olica de Chile, Departamento de 
	      Astronomia y Astrofisica,
	      Casilla 306, Santiago 22, Chile\\	
              email: fcourbin@astro.puc.cl
              \and
	      European Sourthern Observatory, Casilla 19, Santiago, Chile\\
              email: clidman@eso.org
	      \and
	      European Sourthern Observatory, Karl-Schwarzschild Strasse 2,
              D-85748 Garching bei M\"unchen, Germany\\
	      email: gmeylan@eso.org
              \and
              Laboratoire d'Astrophysique, Observatoire Midi-Pyr\'en\'ee, 
              UMR5572, 14 Avenue Edouard Belin, F-31000, Toulouse, France\\
              email: kneib@obs-mip.fr
              \and
	      Institut d'Astrophysique et de G\'eophysique de Li\`ege,
	      Avenue de Cointe 5, B-4000 Li\`ege, Belgium\\
	      email: Pierre.Magain@ulg.ac.be
             }

   \date{Received, 1999; accepted, 1999}

   \titlerunning{Near-IR spectroscopy of HE~1104$-$1805}
   \authorrunning{F. Courbin et al.}	
   \maketitle

   \begin{abstract}
	
A new technique for the spatial deconvolution of spectra is applied to
near-IR  (0.95   -  2.50~$\mu$)   NTT/SOFI  spectra  of   the  lensed,
radio-quiet quasar HE~1104$-$1805.  The  continuum of the
lensing galaxy is revealed  between 1.5~$\mu$ and 2.5~$\mu$.  Although
the spectrum  does not  show strong emission  features, it is  used in
combination  with  previous optical  and  IR  photometry  to infer a
plausible redshift in  the range $0.8<z<1.2$. Modeling of the
system shows  that the lens is  complex, probably composed  of the red
galaxy seen  between the quasar images  and a  more extended component
associated  with a galaxy cluster  with fairly  low velocity
dispersion  ($\sim$   575  ~{km\thinspace  s$^{-1}$}).    Unless  more
constrains  can be put  on the  mass distribution  of the  cluster, 
e.g. from deep  X-ray observations, HE~1104$-$1805
will {\it not} be  a good system to determine H$_0$. 
We stress that  {\sl multiply  imaged  quasars with
known time delays might prove  more useful as tools for detecting dark
mass in distant lenses than for determining cosmological parameters}.

The spectra of the two lensed images of the source are
of great interest.  They show no trace of reddening at the redshift of
the  lens  nor at  the  redshift of  the  source.   This supports  the
hypothesis  of  an  elliptical  lens.   Additionally,  the  difference
between the spectrum  of the brightest component and  that of a scaled
version of  the faintest component is a  featureless continuum.  Broad
and narrow emission lines,  including the FeII features, are perfectly
subtracted.  The very  good quality of our spectrum  makes it possible
to fit  precisely the optical Fe  II feature, taking  into account the
underlying continuum over a wide wavelength range.  HE~1104$-$1805 can
be classified  as a  weak Fe  II emitter.  Finally,  the slope  of the
continuum in the brightest image  is steeper than the continuum in the
faintest image  and supports  the finding by  Wisotzki et  al.  (1993)
that  the  brightest  image  is  microlensed.   This  is  particularly
interesting  in   view  of  the  new  source
reconstruction  methods from multiwavelength  photometric  monitoring.  While
HE~1104-1805   does  not   seem  the   best  target   for  determining
cosmological parameters,  it is  probably the second  most interesting
object   after  Q~2237+0305   (the  Einstein   cross),  in   terms  of
microlensing.


\end{abstract}

%
%

\begin{figure*}[t]
\resizebox{18.0cm}{!}{\includegraphics{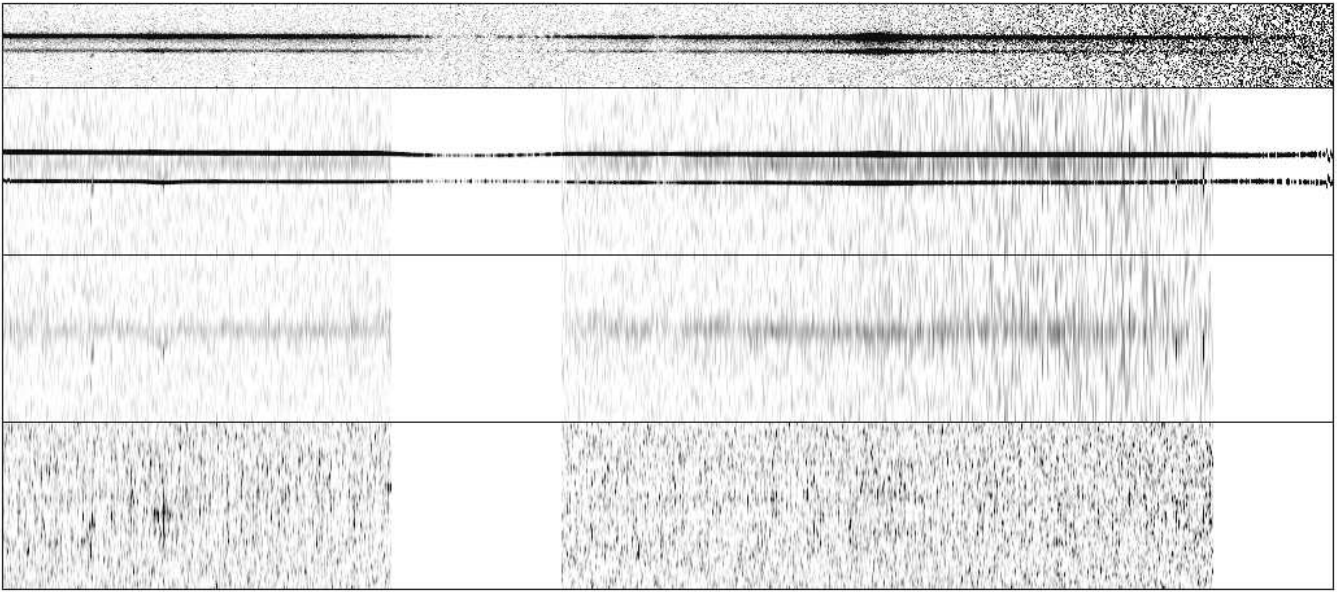}}
\caption[]{ Two dimensional  spectra  of HE~1104$-$1805.  From  top to
bottom, (i) the 2-D  near-IR combined (1.5-2.5~$\mu$) spectrum (seeing
$\sim$   0.6 \arcsec, pixel    size   $\sim$ 0.14\arcsec), (ii)    its
deconvolved  version (resolution of  0.14  \arcsec, pixel size  $\sim$
0.07\arcsec),  (iii) the deconvolved spectrum of   the lens alone, and
(iv) the residual map (see text).}
\end{figure*}

\section{Introduction}

HE~1104$-$1805 is  one object in  the growing list  of gravitationally
lensed  quasars   which  might  be  used   to  constrain  cosmological
parameters.   It was discovered  in the  framework of  the Hamburg/ESO
Quasar Survey and consists of  2 lensed images of a radio-quiet quasar
(RQQ) at $z$= 2.319,  separated by $\sim$3.15\arcsec\ (Wisotzki et al.
1993).  The  lensing galaxy was  discovered from ground  based near-IR
(Courbin,  Lidman \&  Magain,  1998; hereafter  C98)  and HST  optical
observations (Remy et al.   1998; hereafter R98).  The relatively wide
angular  separation  between  the  quasar  images  makes  this  object
suitable for  photometric monitoring programs, as conducted  at ESO by
Wisotzki et  al.  (1998; hereafter  W98).  From light  curves measured
over a period of 6 years, they derived a time delay for HE~1104$-$1805
of $\Delta t = 0.73$ years, with a second possible value of 0.3 years.
Although  we  show that  the  complex  lensing  potential involved  in
HE~1104$-$1805  makes it difficult  to determine  H$_0$ from  the time
delay,  we also  show that  HE~1104$-$1805  is probably  much more  of
interest  for  microlensing studies,  provided  the  lens redshift  is
better known.  The  present paper describes an attempt  to measure the
redshift of  the main lensing  galaxy from near-IR  spectroscopy.  Our
near-IR observations  where motivated by the very  red colors measured
for the lensing galaxy (C98,  R98), and by the better contrast between
the lens and the quasar  in the near-IR. Although we were unsuccessful
in  measuring the  lens redshift  accurately, we  did obtain  high S/N
spectra of the lensed quasar, between 0.95 and 2.5 microns.

%
%

\section{Observations-Reductions}

The data were  taken with SOFI, the near-IR  (1 to  2.5~$\mu$) imaging
spectrograph on the ESO NTT.  Two grisms  were used to  cover the 1 to
2.5~$\mu$ wavelength range: a ``blue grism'',  which covers the region
from 0.95  to 1.64~$\mu$ and a ``red  grism'' which covers  the region
1.53 to 2.52~$\mu$.   With a 1\arcsec\,  slit, the spectral resolution
is around 600.  The observations with the blue grism were taken on the
night of  1998 June 13, for a  total integration time of  4560 seconds
and the  observations with the red  grism were taken  on the  night of
1999   January 6,  with a   total integration  time  of 2400  seconds.
Although the    seeing for  both   nights was   good,   0.6\arcsec\  -
0.8\arcsec, neither night was photometric.

The   slit was  aligned  with the  two images   of  the quasar.  As is
standard practice in the  infrared,  the object  was observed at   two
positions  along the slit.  The  strong and highly  variable night sky
features were effectively removed by subtracting the resulting spectra
from   each  other.   The  2-D     sky-subtracted spectra were    then
flat-fielded, registered, and added.

The  two dimensional  combined  frames were  spatially deconvolved  in
order to extract the spectrum of the lensing galaxy. For this purpose,
we  used the method  outlined by  Courbin et  al.  (1999,  2000).  The
algorithm is  a spectroscopic extension  of the so-called  ``MCS image
deconvolution  algorithm''  (Magain   et  al.   1998).   It  spatially
deconvolves 2-D  spectra of blended  objects, using the spectrum  of a
reference point  source.  It also improves their  spatial sampling and
decomposes them into the individual  spectra of point sources (the two
quasar  images) and extended  sources (the  lensing galaxy).  One also
obtains a  two-dimensional residual map, i.e.,  the difference between
the data and the deconvolved  spectrum (reconvolved by the spectrum of
the  PSF),  in  units  of  the  photon  noise.   The  quality  of  the
deconvolution is checked using the  residual map, which should be flat
with a mean  value of 1.  The different  products of the deconvolution
are shown in Fig.~1 for the spectrum taken with the red grism.

As     the  data  were   obtained   before   we developed  our spectra
deconvolution algorithm, we did not observe  in an optimal way, in the 
sense  that  no  reference spectrum was obtained (a spectrum of a star
in the   field of view).  We   aligned the slit   along the two quasar
components, as is  usually done for  such observations.   However, the
seeing of the data taken with the red  grism was good enough to derive
the  PSF  spectrum  from  the  brighter  quasar  itself. This  was not 
possible with the data taken with the blue grism.

For the  observations taken  with the red  grism, the spectrum  of the
lensing  galaxy,  was  extracted  from the  2-D  deconvolved  spectrum
(extended   component   only)   with  standard   aperture   extraction
techniques.   The quasar spectra  are a  product of  the deconvolution
process and  therefore do  not show any  contamination by  the lensing
galaxy.  For the observations taken  with the blue grism, we used wide
apertures to extract  the quasar spectra from the  original data.  The
lens  is therefore  contaminating the  spectra of  the quasar,  but by
virtue of its very red color, the contamination is negligible.

All extracted 1-D  spectra were then divided by that  of a bright star
and multiplied  by a  blackbody curve that  has a temperature  that is
appropriate for the  spectral type of the star.   Before the division,
spectral features that were visible in the spectra of the bright star,
such  as the Pachen  and Bracket  lines of  hydrogen, were  removed by
interpolation.

%
%

\section{Near-IR spectroscopy of the lens}

\subsection{Plausible lens redshift}

The galaxy spectrum is shown in Fig.~2. The signal-to-noise ratio is
very  low, so  the spectrum  has  been smoothed  with a  box car  with
200~\AA\,  width.  Also plotted  are the  broadband magnitudes  of the
lens (C98, R98 and  Hjorth, private communication).  The lens spectrum
is scaled to match the $H$ and $K$ band magnitudes.

The spectrum  does not lead  to a redshift measurement.   However, the
broadband colours suggest a  significant break in the spectrum between
the $I$ and $J$ bands. We have used the publicly available photometric
redshift code {\it hyperz} (Bolzonella, Miralles \& Pell\'{o} 2000) to
estimate the redshift of the lens.

Since there  is little evidence for  dust in the  quasar spectrum (see
below), we have fitted dust free  models to the data. The best fitting
model  is  a  galaxy  that  was  formed in  a  single  burst  of  star
formation. The best fitting redshift  is $z=1$ with a 1-sigma range of
0.8 to 1.2. The age of the burst is 1.7 Gyrs. The quoted errors on the
redshift do  not include  systematic errors that  could be due  to the
heterogenous nature of  the photometric data, which is  derived from a
mixture of ground and space based observations.

The estimated  redshift is slightly  higher than those  estimated from
the position of the lens  on the fundamental plane ($z=0.77 \pm 0.07$;
Kochanek  et  al.   2000) or  from  lens  models  and the  time  delay
($z=0.79$;  W98).   Note  however   that  mo\-dels  including  a  dark
component (see  next section) can  cope with any redshift  between 0.7
and 1.3  and reproduce the  observed time delay, assuming  for example
H$_0=60$~{km\thinspace s$^{-1}$\thinspace Mpc$^{-1}$}.

As the  break between 0.7 and  1.0~$\mu$ in the model  spectra is very
strong,  a  deep  spectrum in  this  region  is  probably the  key  to
accurately measure its redshift.

\begin{figure}[t]
\leavevmode
\resizebox{8.6cm}{!}{\includegraphics{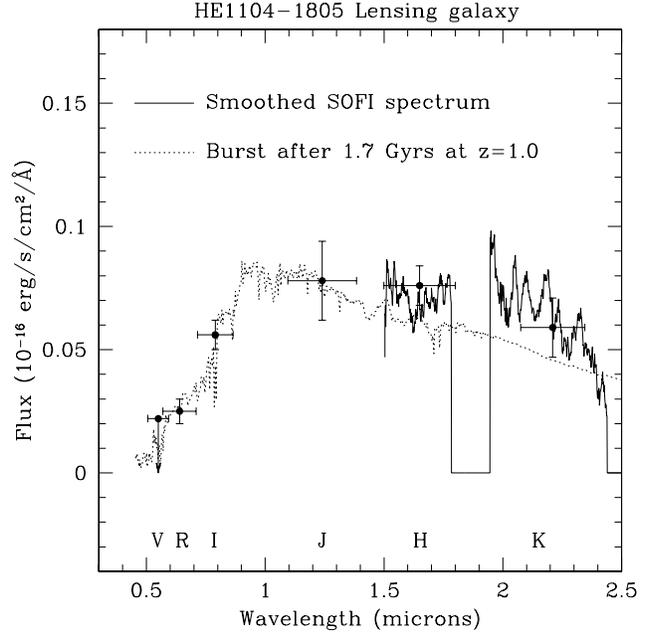}}
\caption[]{Summary of the spectroscopic and photometric data available
for  the  lensing  galaxy  in  HE~1104$-$1805. Also  plotted,  is  the
spectrum of a 1.7 Gyr old burst at z=1.}
\end{figure}

\subsection{Evidence for a high redshift cluster-lens.}

With  only  two  quasar  images  available to  constrain  the  lensing
potential, the unusual image  configuration of HE~1104$-$1805 (R98) is
very difficult  to model uniquely.   The system can not  be reproduced
with a Singular Isothermal  Sphere.  Additional shear and convergence,
whatever their origin may be (intrinsic ellipticity of the lens and/or
intervening  lenses),   are  required  to   match  simultaneously  the
positions and  flux ratio of  the quasar ima\-ges  ($f_A/f_B$=2.8, see
section 4).  In a first model, we introduce an ellipticity in the lens
model,  i.e., we  choose  an isothermal  ellipsoid.   Although we  can
easily  obtain a good  $\chi^2$ fit,  the resulting  model has  a very
large  velocity dispersion, over  300~{km\thinspace s$^{-1}$},  and an
unrealistic   ellipticity  compared  with   the  ellipticity   of  the
associated light  distribution.  Finally,  such a model  predicts time
delays of  $470 h_{50}$  days, while the  observed value is  265 days,
according to W98.

The uncertainty on  the lens redshift can not  explain the discrepancy
between  the measured and  predicted time  delays: additional  mass is
required  to describe  the image  configuration, flux  ratio  and time
delay. We  therefore adopt a  two compo\-nent model including  (1) the
main  lensing   galaxy,  with   ellipticity  and  position   angle  as
constrained by  the light distribution  of the main galaxy  lens, and,
(2) a  more  extended compo\-nent  mimi\-cking  an intervening  galaxy
cluster.  For simplicity,  we centered the cluster on  the main galaxy
and assume an  elliptical isothermal mass profile with  a core radius.
The  fitted   parameters  were  only  the   velocity  dispersion,  the
ellipticity  and  position angle.   Both  the  main  lens and  cluster
components are ta\-ken  to be at redshift 1.0.  Our  best fit model is
shown in  Fig.~3.  It involves a  cluster with moderate  mass, i.e., a
velocity  dispersion   of  $\sigma  \sim   575  \pm  20$~{km\thinspace
s$^{-1}$}.      The     ellipticity     is     0.3     (defined     as
$e=[1-(a/b)^2]/[1+(a/b)^2$])  and PA=10\,  degrees, which  is slightly
tilted  relative to  the  axis of  the  main lens  (which has  PA=46\,
degrees) and with  the light profile of the galaxy  visible in the HST
images  of Lehar et  al.  (1999),  who gives  PA=63$\pm$17\, degrees).
Adding  the cluster  component also  allows  one to  match better  the
observed  shape  parameters of  the  main  lensing  galaxy.  With  the
presence  of the  cluster, the  models can  accommodate a  PA  of 46\,
degrees for the  main lensing galaxy and a  lower velocity dispersion,
$\sigma \sim$ 235~{km\thinspace s$^{-1}$}.

\begin{figure}[t]
\leavevmode
\resizebox{8.6cm}{!}{\includegraphics{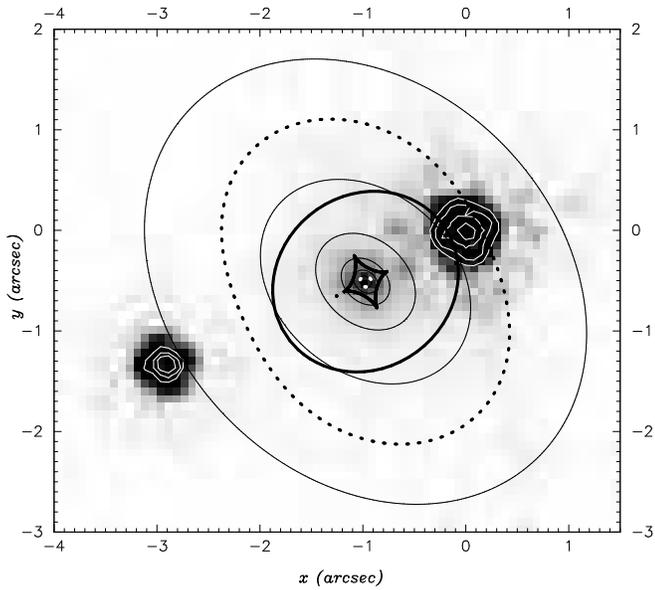}}
\caption[]{HST-NICMOS2 image of HE~1104$-$1805 (PI: Falco) obtained in
the $H$-band,  and lens model. The  thin solid lines  show the isomass
density contours for our galaxy$+$cluster lens model. The thick dashed
lines  are the  critical  curves and  the  thick solid  lines are  the
caustic lines at the redshift of  the quasar. The dot left to the diamond
shape caustic shows the source position for our best fit model.}
\end{figure}
\begin{figure}[t]
\resizebox{8.7cm}{!}{\includegraphics{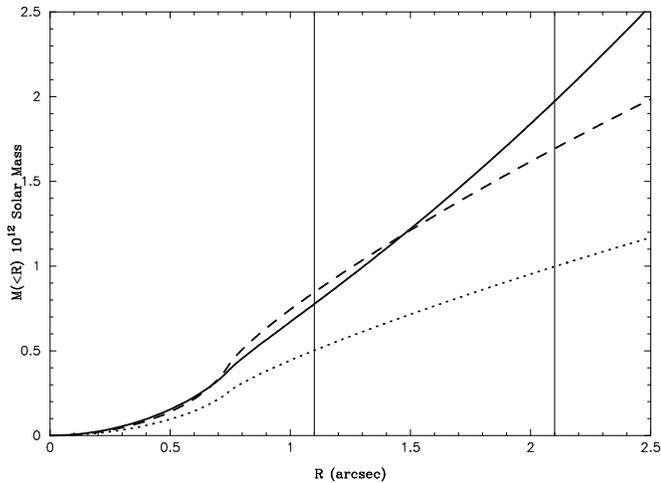}}
\caption[]{Total mass  within a surface  of radius\, $r$ for  the single
galaxy lens model (long dash), galaxy$+$cluster model (solid), and for
the galaxy  component of the galaxy$+$cluster model  (short dash). The
positions of the quasar images are marked with vertical lines.}
\end{figure}

If  we assume  H$_0=60$~{km\thinspace  s$^{-1}$\thinspace Mpc$^{-1}$},
$\Omega=0.3$ and $\Lambda=0.7$,  the galaxy$+$cluster model reproduces
well the  measured time delay, giving  a value of $\Delta  t \sim 265$
days.   However, we  stress that  the lens  redshift and  the velocity
dispersion  of the  cluster are  redundant parameters:  increasing the
cluster's mass or decreasing the lens redshift have the same effect on
the  time delay.   This  degeneracy between  the  two parameters  will
prevent any estimate of H$_0$ until more observational constraints are
available on  the cluster component  of the lensing matter.   The time
delay now available  in HE~1104$-$1805 can therefore be  seen as a new
important constraint on  the lens model, indicating the  presence of a
yet undetected  cluster, rather  than a way  to constrain  H$_0$.  The
mass within an area of a given radius is shown in Fig.~4 for different
model components.   

If real, the  cluster we predict in our model  is difficult to detect,
with  only  $\sigma  \sim  575 \pm  20$~{km\thinspace  s$^{-1}$}.   At
a redshift  of 1, it  would be  even more  difficult to  see than  the more
massive clusters involved in  other lenses such as AX~J2019+112 (e.g.,
Benitez et al.  1999) or  RX~J0911.4+0551 (Burud et al. 1998, Kneib et
al.  2000).   However the velocity  dispersion of such a  cluster will
change depending on the cluster  center position. If it is not aligned
with  the main  galaxy  lens, its  velocity  dispersion will  increase
significantly. Deep  X-ray observations and/or deep IR  images of this
field would be invaluable in constraining further the models.

%
%

\section{Near-IR spectroscopy of the source at $z$~=~2.319}

The  0.95  -  2.50~$\mu$ spectrum  of  the  quasar  pair is  shown  in
Fig.~5. The spectra  are on a relative flux  scale.  Regions of high
atmospheric absorption are set to  zero.  The spectra show clearly the
Balmer lines: H$\alpha$, H$\beta$,  H$\gamma$ and a partially obscured
H$\delta$, the [OIII] doublet and several broad FeII features (Francis
et al.  1991).   From the Balmer lines, the redshift  is 2.323 for the
brighter component (component A)  and 2.321 for the fainter (component
B). The measurement error is $\Delta  $z$ = 0.002$, so the redshift of
the two components agree with each other, but are slightly larger than
the  determination  at   optical  wavelengths  ($z=2.317$,  Smette  et
al. 1995).  As with most  quasars (McIntosh et al.  1999b), the [OIII]
doublet  is slightly  blue  shifted ($z=2.319$)  with  respect to  the
Balmer lines.

\begin{figure}[t]
\resizebox{9.0cm}{!}{\includegraphics{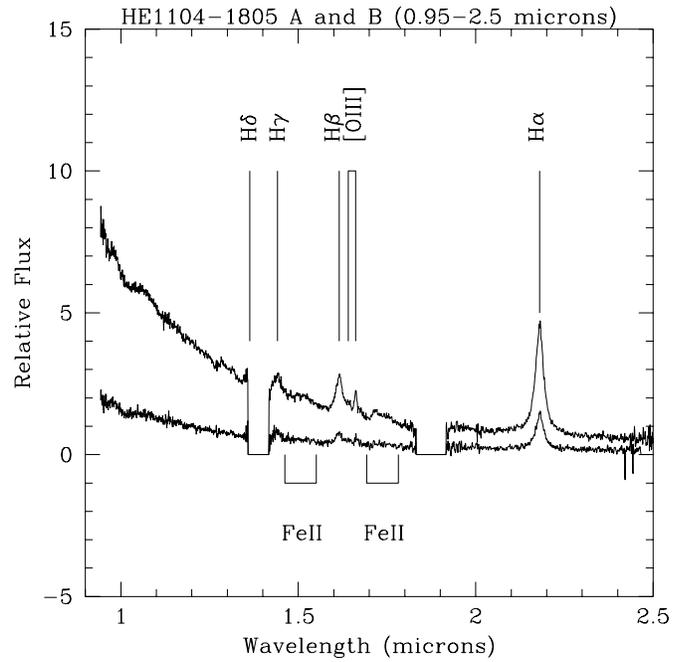}}
\caption[]{One     dimensional   near-IR  spectra   of
components A and B of HE~1104$-$1805.}
\end{figure}

Following Wisotzki  et al. (1993),  we subtracted a scaled  version of
the   fainter    component   from   the   brighter    one,   that   is
$f_{\lambda}(A)-c.f_{\lambda}(B)$. The scale is set so that the Balmer
lines  vanish.  We  find that  we  require $c=2.9\pm0.1$  for the  red
spectrum  and $c  = 3.0\pm0.1$  for  the blue  spectrum.  Wisotzki  et
al.  and Smette  et  al.  (1995)  have  used $c  =  2.8$.  The  slight
difference  between   Wisotzki's value and  ours   may  only  reflect
systematic differences in the way  the object was observed and the way
the data were reduced rather  than anything real.  For example, the IR
observations were done with a one-arc-second slit, and any small error
in the alignment angle could cause such a difference.

The difference spectra are plotted in Fig.~6.   Here we plot the raw
difference spectra as  the dotted line, and a  smoothed version of this as
the continuous  line.  The spectrum  of the brighter component is also
displayed.   The difference   spectra are featureless.   The  residual
after subtracting the strong  H$\alpha$ line  is less  than  1\%.  Not
only are the broad hydrogen features removed from the spectra, but the
broad iron features and  the [OIII] doublet  are  removed as well.  As
noted by Wisotzki et al.  (1993) there  appears to be excess continuum
in the brighter component.

\begin{figure}[t]
\resizebox{9.0cm}{!}{\includegraphics{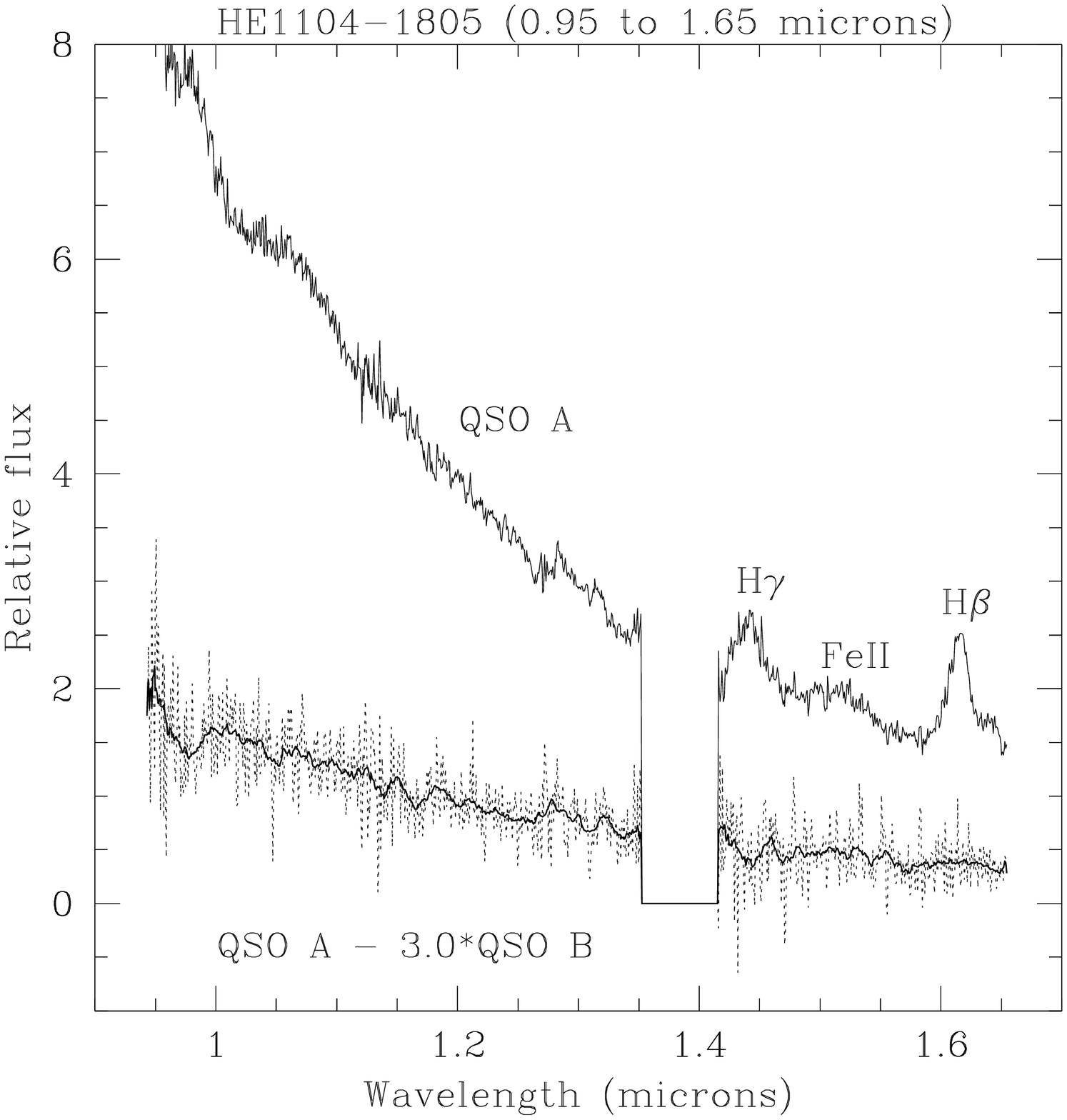}}
\resizebox{9.0cm}{!}{\includegraphics{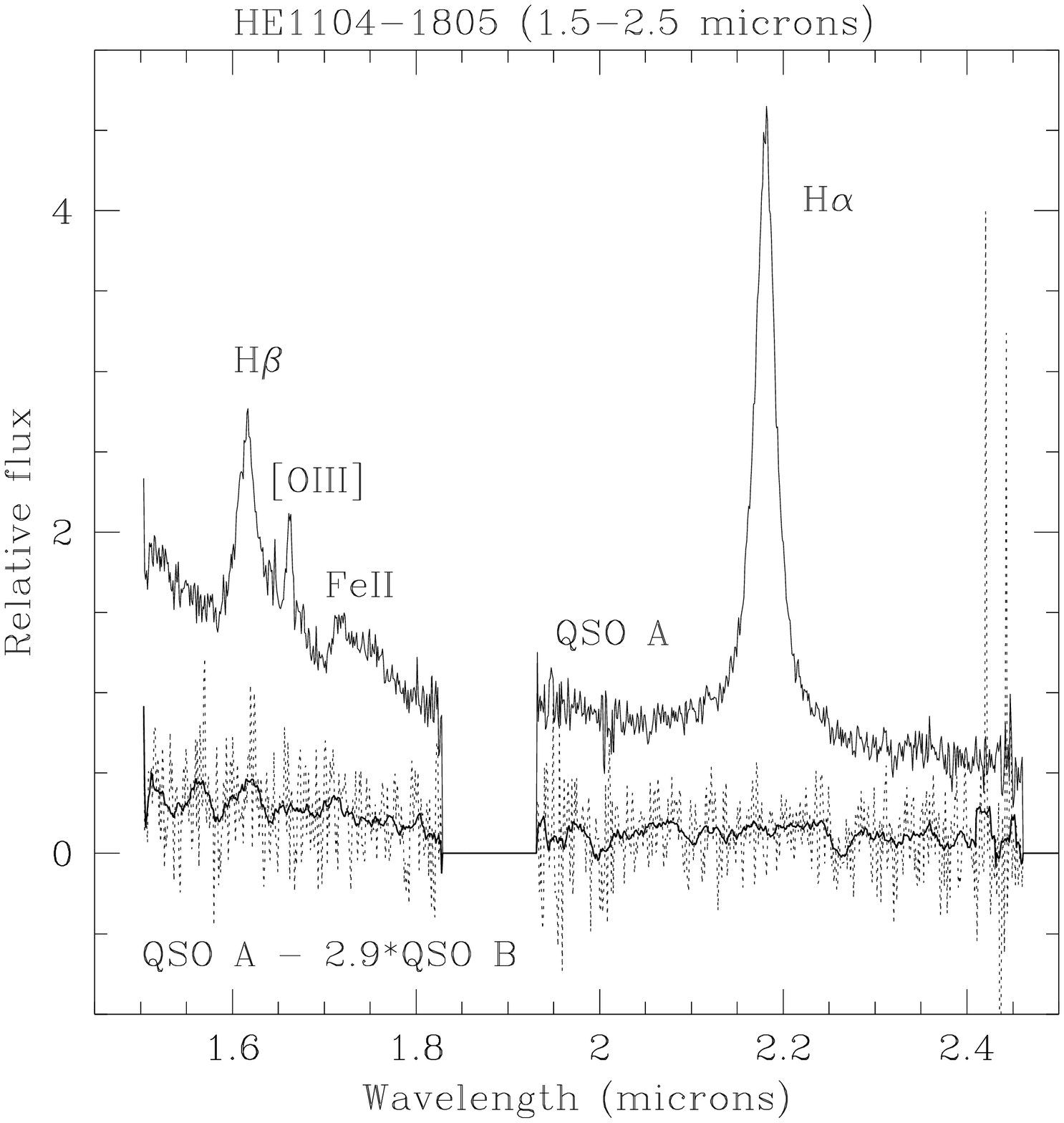}}
\caption[]{Difference spectra of the   two quasar images for the  blue
grism (top) and  red grism (bottom).   The raw difference spectra  are
plotted   as the dotted  lines,  a  smoothed  version of these  as the
continuous  lines.  The spectra   of the  brighter component  is  also
displayed.}
\end{figure}

%
%

\subsection{Extinction}

The Balmer decrement is around 4 for both components, and this is well
within the  range expected for unreddened quasars  (e.g., Baker et al.
1994).  Thus, there  is no evidence  for absolute reddening.  However,
the limits  we can set on this are weak as the range of values for the 
Balmer decrement in quasars is rather broad.

The limits for differential  reddening are considerably stronger.  The
ratio of the emission lines in the brighter  and fainter components is
$2.9 \pm 0.1$. The error brackets the measured variation of this ratio
over time (six years of observations) and  over wavelength.  It is not
clear if this  variation is real  or the result  of measurement error.
This ratio is remarkably constant  over a large wavelength range, from
CIV  at 1549~\AA\ to  H$\alpha$, and we can used  it to place an upper
bound  on  the  amount of  differential  extinction   between  the two
components. If  we assume that  the lens is at  $z=1$ and if we assume
that the standard galactic extinction law (Mathis 1990) is applicable,
then the differential extinction between the two components is $\Delta
E(B-V) < 0.01$ magnitudes.

Recently, Falco  et al.  (1999) measure  a  differential extinction of
$\Delta E(B-V)=0.07\pm 0.01$ for  HE~1104$-$1805 in the sense that the
B component has a higher  extinction. However, their measurements rely
on broad  band    photometry and  their results  can   be mimicked  by
chromatic amplification of the continuum region by microlensing. If we
were  to repeat the experiment by   comparing the relative strength of
the continuum   at 1.25~$\mu$  and  2.15~$\mu$,   we would  derive   a
differential extinction  of   $\Delta E(B-V)=0.16$  magnitudes  and we
would find also that B component was differentially reddened.

%
%

\subsection{Emission line properties of the source.}

The  emission  line properties  of  high  redshift  quasars have  been
examined  for  correlations  between  line ratios  and  equi\-va\-lent
widths  (McIntosh et  al.  1999a,b;  Muramaya  et al.   1999). As  the
signal-to-noise  ratio  and  spectral  coverage  of our  IR  data  are
considerably better, we have  re-measured the emission line parameters
for HE~1104$-$1805.

\begin{figure*}[t]
\resizebox{9.0cm}{!}{\includegraphics{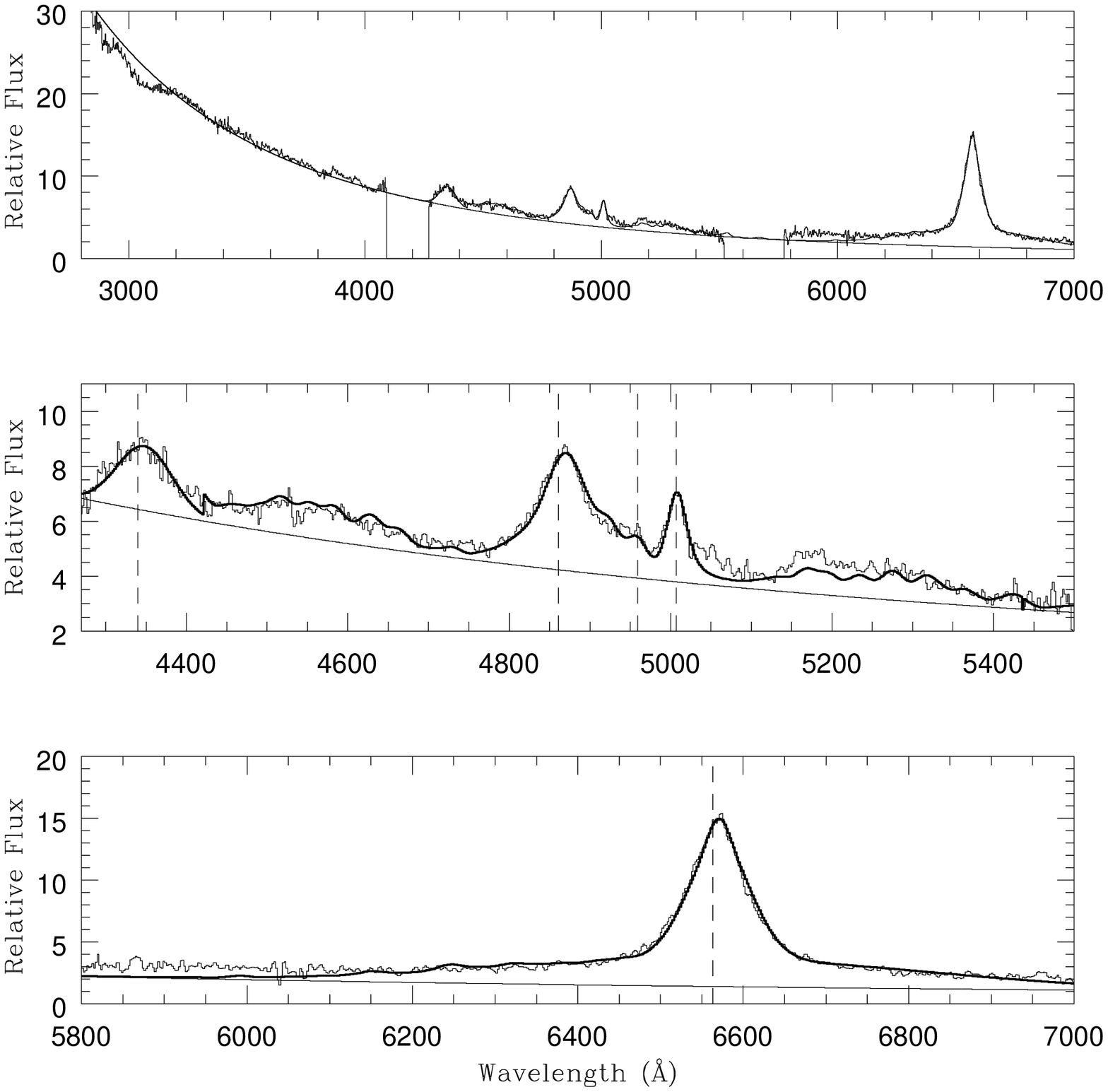}}
\resizebox{9.0cm}{!}{\includegraphics{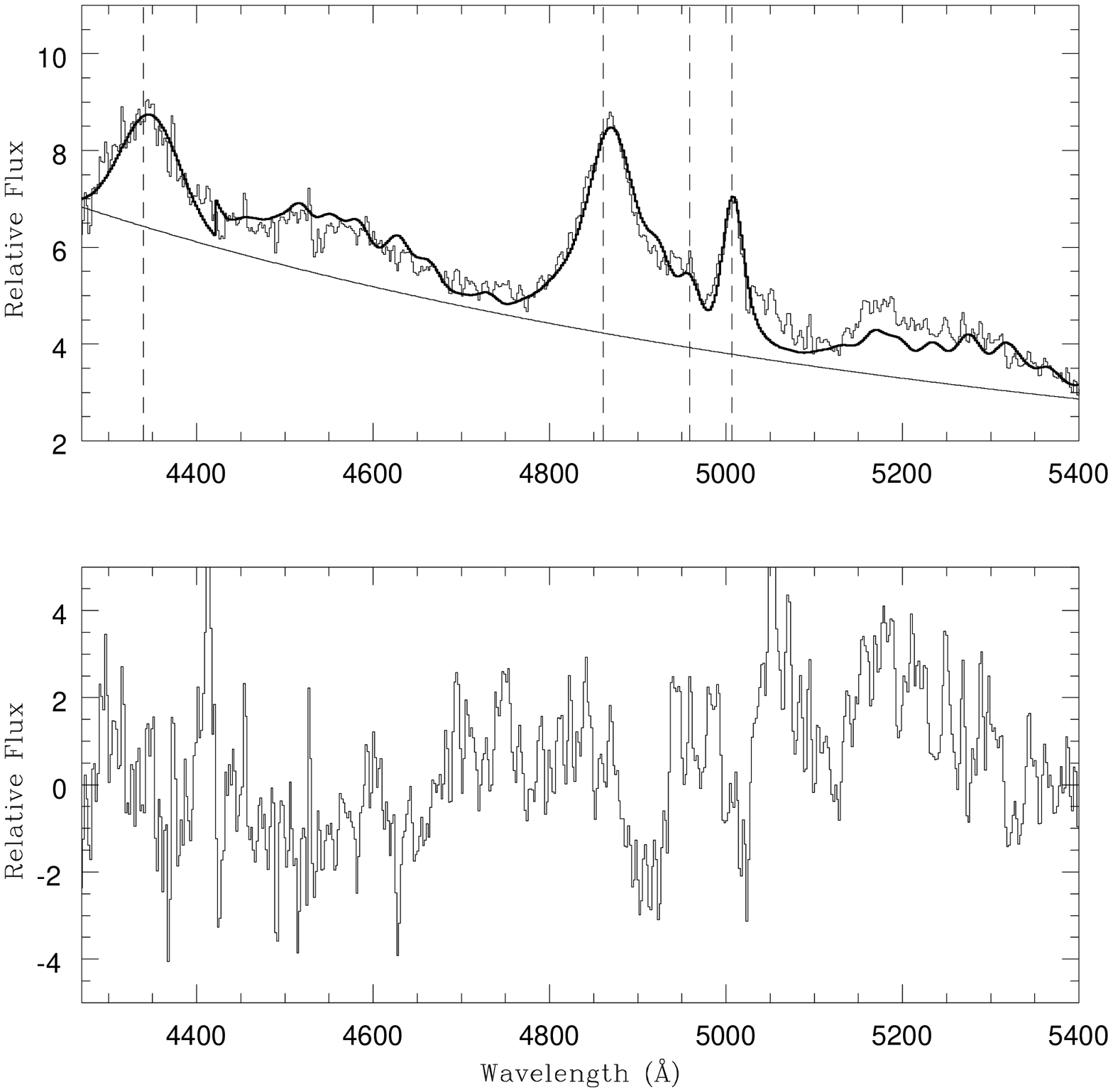}}
\caption[]{{\bf  Left:} The  top  left panel  displays the  rest-frame
spectrum of component  A and its fit as described in  the text.  It is
the  entire  spectrum (multiplied  by  $1+z_{syst}$),  with the  model
spectrum superposed  as a  bold line.  The  lower two left  panels are
both  zooms in  the  left and  right  halves of  the  top panel.   The
vertical dashed lines  are drawn at the rest-frame  wavelength of each
emission     line,    considering     a    systemic     redshift    of
$z_{syst}=2.319$.~{\bf  Right:} A  zoom  in the  optical FeII  region,
again with  the fitted spectrum  superposed.  The bottom panel  is the
difference  between the  data  and the  fit,  in units  of the  photon
noise.}
\end{figure*}

Fitting of the spectrum  was  done in a  similar  way to that  used in
McIntosh et al.   (1999a), but with extended   spectral coverage.  The
model spectrum is a sum of Gaussian lines superposed on an exponential
continuum  to  which is  added   a  numerical optical  FeII   template
(4250~\AA\, and 7000~\AA).  The iron template  consists of the optical
spectrum   of I~Zw~1  obtained by   Boroson  \& Green  (1992).  Before
computing  the  model  spectrum  the    template  is smoothed to   the
resolution of our observations by  convolving it with a Gaussian  line
which  has the same  FWHM than the broad  emission lines of the quasar
(CIV,  in  the  present   case), i.e.,  a  rest-frame   width of  6400
{km\thinspace  s$^{-1}$},   or  14~\AA.    A  systemic   redshift   of
$z_{syst}=2.319$ is determined  from the [OIII] $\lambda$5007 emission
line   and  applied to  the  data   to  obtain a  rest-frame  spectrum
(multiplied by  $1+z_{syst}$ to conserve  flux).  As  the positions of
all other emission lines are redshifted relative to the [OIII] line by
different  amounts, their  wavelengths  are adjusted  independently of
each other.  We   measure  a mean  redshift of  $z_{Balmer}=2.323  \pm
0.001$ from the H$\gamma$  $\lambda$4340, H$\beta$ $\lambda$4861,  and
H$\alpha$ $\lambda$6562 emission lines.  We used a sum of Gaussians to
fit each Balmer line.  This  arbitrary decomposition is certainly  not
aimed at  being representative of any physical  model but still allows
us  to measure   fluxes.  One single  Gaussian was   used to  fit  the
H$\gamma$ line  while two Gaussians are required  to fit  H$\beta$ and
three to fit H$\alpha$ which  shows very wide  symmetrical wings.  The
[OIII] doublet is represented by two Gaussians with a fixed line ratio
of three (between [OIII] $\lambda$4959 and  [OIII] $\lambda$5007). All
line widths are fixed during the fit and  all intensities are adjusted
simultaneously (with   the  conjugate  gradient  algorithm) with   the
strength of the iron template and  exponential continuum.  The results
of the  fit are   reported in Table~1   and  Fig.~7.  The best   fit
spectrum has  a    power law  continuum  of  the    type  $F_\lambda =
\lambda^{-\alpha}$,  with   $\alpha=3.6$.     One-sigma    errors were
estimated by running the fit with different  line widths.  In addition
to these  errors, one  should consider  the  error introduced   by the
continuum   determination.  Changing the    index of  the  exponential
continuum by 10\% can affect iron flux measurement by up to 20\%.  The
other, much narrower emission lines,  are less affected, but we stress
the need  for continuum fitting over  a very wide  wavelength range in
order  to minimize such  systematic errors.   This  was pointed out by
Murayama et al.  (1999).  It is now obvious on our better data.

\begin{table*}[t]
\caption{Rest-frame emission line properties of HE~1104$-$1805, as measured
from the fit performed in section 4.2.}
\begin{tabular}{l c c c c}
\hline \hline 
     & Flux & FWHM  & FWHM &  Eq. Width \\ 
     & (Arbitrary units)  & (\AA) & {km\thinspace s$^{-1}$} &  (\AA) \\ 
\hline
FeII (4434-4685) & 1.026 $\pm$ 0.003 & $-$ & $-$ & 18.73 $\pm$ 0.07\\
FeII (4810-5090) & 0.608 $\pm$ 0.002 & $-$ & $-$ & 14.92 $\pm$ 0.05\\
\hline
H$\beta$ (narrow)& 1.283 $\pm$ 0.080 & 50 $\pm$ 3  & 3084 $\pm$ 180 & $-$\\
H$\beta$ (broad) & 1.922 $\pm$ 0.080 & 120 $\pm$ 10& 7404 $\pm$ 620& $-$\\
H$\beta$ (total) & 3.205 $\pm$ 0.120  &    $-$     &  $-$  & 74 $\pm$ 1 \\
\hline
H$\gamma$    & 1.860 $\pm$ 0.010 & 80 $\pm$ 3  & 5528 $\pm$ 210& 28 $\pm$ 2 \\
\hline
H$\alpha$ (narrow) & 0.747 $\pm$ 0.200 & 30 $\pm$ 3 & 1370 $\pm$ 140 & $-$ \\  
H$\alpha$ (medium) & 8.739 $\pm$ 0.030 & 90 $\pm$ 3 & 4110 $\pm$ 140 & $-$ \\
H$\alpha$ (broad)  &13.49 $\pm$ 0.030 & 600 $\pm$ 30& 27400 $\pm$ 1400 & $-$ \\
H$\alpha$ (total)  & 22.97 $\pm$ 0.210 & $-$        & $-$ & 1563 $\pm$ 15 \\
\hline
OIII (4959) & 0.194 $\pm$ 0.010 & 25 $\pm$ 2 &1512 $\pm$ 120 & $-$ \\
OIII (5077) & 0.581 $\pm$ 0.030 & 25 $\pm$ 2 &1496 $\pm$ 120 & 14.8 $\pm$ 0.7\\
\hline
\end{tabular}
\end{table*}

The quality of the fit  is overall very good;  however, there are some
regions  where significant differences   exist,  as indicated by   the
residuals shown in the  bottom left panel  of Fig.~7.  Most  notably
the FeII complex  red-wards of [OIII]  is relatively stronger that the
FeII complex blue-wards of H$\beta$.

The ratio of  the EWs of [FeII] to H$\beta$ is  0.20. This is slightly
lower  than that  measured by  McIntosh  et al.   (1999a), who  report
0.29$^{+0.07}_{-0.09}$. The difference is probably not significant but
there  are  two  systematic  biases  that  make  a  direct  comparison
difficult.   Firstly, the continuum  in this  fit is  well determined,
whereas the small spectral coverage of the previous work may mean that
EWs are  underestimated (Muramaya et  al.  1999).  Secondly,  and more
fundamentally,  EWs  measured in  lensed  quasars  are susceptible  to
microlensing which preferentially  amplifies the continuum rather than
the larger emission line regions.  In fact, the continuum for unlensed
quasars also varies.   This means that EWs are a  poor measure to use.
Line fluxes are not susceptible to continuum variations, no matter how
the continuum varies, whether it is  intrinsic to the AGN or caused by
microlensing.        From        our       fit,       we       measure
F(FeII(4810-5090))~/~F(H$\beta$)~=~0.18~$\pm$~0.04                  and
F(FeII(4434-4685))~/~F(H$\beta$)~=~0.32~$\pm$~0.04 which, according to
Lipari  et al.   (1993) makes  of  HE~1104$-$1805 a  rather weak  FeII
emitter.

%
%

\section{Is microlensing detected in HE~1104$-$1805~?}

The  possibility  of microlensing  can  be  judged  from a  comparison
between  the size  of  the Einstein  ring  and the  size  of a  quasar
continuum emitting region.  The latter is thought to be produced by an
accretion disk and is of the order of $10^{14}$ to $3\times10^{15}$ cm
(Wambsganss  et al.  1990; Krolik  1999 and  references  therein). The
former depends on the mass  of the microlenses, $M$, and is $3.0\times
10^{16} \sqrt(M/M_{\odot})$  cm.  In  this calculation and  those that
follow, we have assumed that the lens is at $z=1$, and we have assumed
a    cosmology    where   $H_0=60$~{km\thinspace    s$^{-1}$\thinspace
Mpc$^{-1}$},  and  $(\Omega_M,\Omega_{\Lambda})  = (0.3,0.0)$.   Thus,
given  that  there  is  a  suitable  alignment,  microlensing  of  the
continuum is  possible.  Furthermore, the spectrum of  the A component
is considerably harder than that of the B compo\-nent.  This chromatic
effect  supports the  idea  that the  continuum  is microlensed  since
higher energy photons come from  the inner part of the accretion disk,
and are hence more susceptible to high amplification microlensing than
lower energy photons.  In fact, with suitable modeling of the lens and
additional spectroscopic data, it may be possible to place constraints
on accretion disc models (e.g., Agol \& Krolik, 1999).

The  likelihood  of  microlensing  then  depends  on  the  density  of
micro-lenses. If we model the  mass distribution of the lensing galaxy
as in  section $3.2$, we  can use the  distance between the  two macro
images to  determine the  mass density at  each image  position.  This
gravitational convergence  or optical  depth, $\kappa$, is  quite high
for both  components.  For the  A component, it is  $\kappa=0.73$; for
the B  component, it is 0.53. If  this is made entirely  of stars then
microlensing  of  either  component  is  highly  likely.   In  detail,
however, only the main lensing galaxy is contributing to microlensing.
The  actual  microlensing  optical  depth  at  the  two  quasar  image
positions is then lower, but still high.

The  typical time-scale  between two  consecutive  microlensing events
depends  on the  transverse velocity  of the  source and  the velocity
dispersion of the microlenses. The velocity dispersion for the lensing
galaxy  is high  (using  a  single galaxy  model  one derives  $\sigma
\approx  300$~{km\thinspace   s$^{-1}$},  or  about  235~{km\thinspace
s$^{-1}$} if  a cluster  is also involved)  and it is  probably larger
than the transverse bulk velocity.  Dividing this velocity directly by
the  diameter of  the Einstein  ring, one  derives a  time scale  of 3
years. This  is quite  long; however, it  has been shown  that stellar
proper  motions  produce  a  higher  microlensing rate  than  the  one
produced  by a  bulk velocity  of the  same magnitude  (Wambsganss and
Kundic 1995, Wyithe et  al.  2000).  Furthermore, the typical duration
of a microlensing event is  the time for the continuum emitting region
($10^{15}\rm  cm$) to cross  a caustic  with velocity  $\sigma \approx
300$~{km\thinspace s$^{-1}$}.   This is of  the order of a  few months
and  much  shorter  that  the time  between  consecutive  microlensing
events.

Thus, it  is likely that  microlensing affects the A  compo\-nent.  As
the stellar density of the  lens near the B component is approximately
half that  of the  A component, it  is quite likely  that microlensing
affects the B  component as well. We should  expect that the continuum
of the A component should be preferentially amplified relative to that
of the  B component for the majority  of the time, but  we should also
expect that the  B component should be preferentially  amplified for a
fraction  of   the  time.   HE~1104$-$1805  has   now  been  monitored
spectroscopically for six years  (Wisotzki et al.  1998).  During that
time, the  continuum of  both components have  been observed  to vary;
however,  the continuum  of the  A  component has  always been  harder
(Wisotzki, private communication).  As  the time delay between the two
components is  of the order  of 0.73 year  (W98), the hardness  of the
continuum  in the  A  component  cannot be  attributed  to time  delay
effects.  The most natural explanation is microlensing.

Additionally, the relative  level of the continuum of  the A component
is  more  variable  than that  of  the  B  component  (see Fig.  2  in
W98). This cannot  be attributed to photometric errors,  because the A
component is a  factor of 3 brighter than B,  both components are well
separated  and  the  lensing   galaxy  is  much  fainter  than  either
component.

Conversely,   the   BLR   does   not   appear  to   be   affected   by
microlensing. From 1993 to 1999,  the ratio of the broad lines between
the two components has varied little, with $2.9~\pm~0.1$ (W98 and this
paper). 

The lines  of the BLR in  the IR spectra presented  here subtract very
cleanly, better than  1\% of the original line  flux. Naively, one may
then expect that any substructure  in the BLR needs to be considerably
larger  than the  microlensing caustics,  i.e., $3\times  10^{16}$ cm.
However,  a  more secure  estimate  requires  better  modeling of  how
microlensing in this  particular lens can affect the  profile of lines
from the BLR (e.g., Schneider \& Wambsganss, 1990).

%
%

\section{Conclusions}

We have obtained 1~$\mu$ - 2.5~$\mu$ spectra of the gravitational lens
HE~1104$-$1805. Although we were not successful in measuring a precise
redshift for the lens, the lens  is probably an early type galaxy with
a plausible redshift of $0.8 <  z < 1.2$. This is slightly larger than
estimates based  on the measured  time delay and estimates  based from
the position  of the lens on  the fundamental plane.  We show however,
that we can reconcile time delay and lens redshift by adding a cluster
component to the lens models.

We  find that  the continuum  in the  A component  is harder  than the
continuum in  the B component.  The most probable explanation  is that
the A component is microlensed by compact objects in the lens galaxy.

The ratio of the emission  lines between the two components is $2.9\pm
0.1$. This  is consistent with  that measured at  optical wavelengths.
The  constancy of  this ratio  over  a large  wavelength range  limits
strongly  the  amount  of  differential  extinction  between  the  two
components.   We  find that  the  differential  extinction is  $\Delta
E(B-V) < 0.01$ magnitudes.

We find that broad and narrow  emission lines can be removed very well
by subtracting  a scaled version of  the spectrum of  component B from
the spectrum of  component A. The residual near  the $H\alpha$ line is
less than $1\%$  of the original line flux. It may  be possible to use
this  near  perfect subtraction  to  limit  models  of the  BLR.  This
possibility should be investigated further.

Finally, we note that the time delay measured in HE~1104$-$1805 allows
us to demonstrate that the lensing potential is composed of a main
lensing galaxy and a more extended ``cluster'' component.  Without the
time delay a single galaxy lens would also have been a viable model.
With the rapidly increasing number of lenses with known time delay, we
can therefore expect to constrain the content in dark matter of lens
galaxies and it may be found that intervening clusters are a lot more
frequent than first thought.

\begin{acknowledgements}
We would like to thank Daniel Mc Intosh and Todd Boroson for providing
us  with the FeII  template used  for the  line fitting.   F.  Courbin
acknowledges financial support through Chilean grant FONDECYT/3990024.
Additional support from the  European Southern Observatory and through
a CNRS/CONICYT grant is  also gratefully acknowledged. Jean-Paul Kneib
acknowledges support from CNRS.
\end{acknowledgements}


\begin{thebibliography}{}

   \bibitem[1999]{ag99} Agol E., Krolik J., 1999, ApJ 524, 49

   \bibitem[1994]{Baker94}  Baker  A.   C.,  Carswell  R.  F.,  Bailey
   J. A. et al. 1994, MNRAS 270, 575

   \bibitem[1999]{Beni99}  Benitez N.,  Broadhurst T.,  Rosati  P., et
   al. 1999, ApJ 527, 31

   \bibitem[2000]{Bolz00} Bolzonella M., Miralles J. M., Pell\'{o} R.,
   2000, astro-ph/0003380

   \bibitem[1992]{Boro92} Boroson T.A., Green R.F., 
   1992, ApJS 80, 109

   \bibitem[1998]{bu98} Burud I., Courbin F., Lidman C., et al.  1998,
   ApJ 501, L5

   \bibitem[1998]{Courbin98a} Courbin F., Lidman C., Magain P.,
   1998, A\&A 330, 57

   \bibitem[1999]{Courbin99} Courbin F., Magain P., Sohy S.,
   Lidman C., Meylan G., 1999, ESO-Messenger 97, 26

   \bibitem[2000]{Courbin20}  Courbin F.,  Magain  P., Kirkove  M.,
   Sohy S., 2000, ApJ 529, 1136

   \bibitem[1999]{Falco99} Falco  E.E., Impey C.D.,  Kochanek C.S., et
   al.  1999, ApJ 523, 617

   \bibitem[1991]{Fran91} Francis  P. J., Hewett P. C.,  Foltz, et al.
   1991, ApJ 373 465

   \bibitem[1999]{kro99} Krolik  J., 1999 ``Active  galactic nuclei :
   from  the  central  black   hole  to  the  galactic  environment'',
   Princeton  Series  in  Astrophysics,  Princeton  University  Press,
   ed. J., P., Ostriker.

   \bibitem[2000]{Koe20} Kochanek  C. S, Falco  E.E., Impey C.  D., et
   al.  2000, ApJ in press

   \bibitem[2000]{Kneib2000}  Kneib  J.-P.,  Cohen J.,  Hjorth  J.,
   2000, ApJ submitted

   \bibitem[1999]{Lehar99}  Lehar  J.,   Falco  E.,  Kochanek  C.,  et
   al. 1999, astro-ph/9909072

   \bibitem[1993]{Lip93} Lipari S., Terlevisch R., Macchetto F.,
   1993, ApJ 406, 451 

   \bibitem[1998]{Magain98} Magain P., Courbin F., Sohy S.,
   1998, ApJ 494, 472

   \bibitem[1990]{Mathis90} Mathis J. S., 1990, ARA\&A 28, 37

   \bibitem[1999a]{McI99a} McIntosh D. H., Rieke M. J., Rix H.-W., Foltz
   C. B., Weymann R. J., 1999a, ApJ 514, 40

   \bibitem[1999b]{McI99b} McIntosh D. H., Rix H.-W., Rieke M. J., Foltz
   C. B., 1999b, ApJ 517, L73

   \bibitem[1999]{mura99}  Murayama  T., Taniguchi  Y.,  Evans A.,  et
    al. 1999, AJ 117, 1645

   \bibitem[1998]{Remy98}  Remy  M., Claeskens  J.-F.,  Surdej J.,  et
    al. 1998, NewA 3, 379

   \bibitem[1995]{Smette95} Smette A., Robertson  J. G., Shaver P. A.,
   et al. 1995, A\&AS 113, 199

   \bibitem[1990]{schnei90} Schneider P., Wambsganss J., 
   1990, A\&A 237, 42

   \bibitem[1995]{Wam93} Wambsganss J., Kundic T., 1995, ApJ 450, 19

   \bibitem[1990]{Wam90} Wambsganss J., Schneider P., Paczynski B., 1990
   ApJ 358, L33

   \bibitem[1993]{Wizot93} Wisotzki L., Koehler T., Kayser R.,
   Reimers D., 1993, A\&A 278, L15

   \bibitem[1998]{Wizot98} Wisotzki L., Wucknitz O., Lopez S.,
   S{\o}rensen A., 1998, A\&A 339, L73

   \bibitem[1999]{Wyi98} Wyithe J. S. B., Webster R., Turner E., 2000,
   MNRAS 312, 843

\end{thebibliography}
\end{document}